\documentclass[referee]{cjaa}           

\usepackage{graphicx}                   
\input{epsf.sty}                        
\input{psfig.sty}                       

\begin{document}

   \title{A GHz Flare in a Quiescent Black Hole and A Determination of the Mass Accretion Rate
}

   \volnopage{Vol.0 (200x) No.0, 000--000}      
   \setcounter{page}{1}          

   \author{Sabyasachi Pal
      \inst{1,3}\mailto{}
   \and Sandip K. Chakrabarti
      \inst{1,2}
      }
   \offprints{Sabyasachi Pal}                   

   \institute{Centre for Space Physics, Chalantika 43, Garia Station Road, Kolkata 700084, India\\
             \email{space\_phys@vsnl.net}
        \and
             S. N. Bose National Centre for Basic science, Salt Lake, Kolkata, 700098, India\\
        \and
             Jadavpur University, Kolkata 700032, India\\}

   \date{Received~~2004 July 25; accepted~~2004~~month day}

\abstract{If the total energy of a radio flare is known, one can estimate the 
mass accretion rate of the disk by assuming equipartition of magnetic 
energy and the gravitational potential energy of accreting matter. We present
here an example of how such an estimate could be done.
Our recent radio observation using the Giant Meter Radio Telescope 
(GMRT) of the galactic black hole transient  A0620-00 at 1.280 GHz 
revealed a micro-flare of a few milli-Jansky. Assuming a black 
hole mass of $10 M_\odot$ for the compact object, we find the accretion rate
to be at the most ${\dot M} = (8.5 \pm 1.4) \times 10^{-11} (\frac{x}{3})^{5/2}
M_\odot$ yr$^{-1}$, where, $x$ is the distance from the hole in units of
Schwarzschild radius. This is consistent with the earlier estimate of the accretion
rate based on optical and X-ray observations. We claim that this procedure
is general enough to be used for any black hole candidate.
}

   \authorrunning{Sabyasachi Pal \& S. K. Chakrabarti}            
   \titlerunning{Accretion rate of A0620-00 in Quiescence}  

   \maketitle

\noindent To be Published in the proceedings of the 5th Microquasar Conference: 
Chinese Journal of Astronomy and Astrophysics


%
%
\section{Introduction}           
\label{sect:intro}

Our understanding of the accretion processes at low accretion rates suggests that 
magnetic field may be entangled with hot ions at virial temperatures 
and could be sheared and amplified to the local equipartition
value (Rees 1984). If so, dissipation of this field, albeit small, should produce
micro-flares from time to time, and they could be detectable especially 
if the object is located nearby. In the case of AGNs and QSOs, the flares are common
and the energy release could carry information about the accretion rates in those
systems. We present here an application of this understanding of the accretion process
in the context of the galactic black hole transient A0620-00 (Pal \& Chakrabarti, 2004).

A0620-00 was discovered in 1975 through
the Ariel V sky survey (Elvis et al. 1975). It is located at a 
distance of $D=1.05$ kpc (Shahbaz, Naylor \& Charles1994).
A0620-00 is in a binary system and its mass is estimated to be 
around $10M_\odot$ (Gelino, Harrison \& Orosz 2001). A0620-00 is not particularly well
known for its activity in radio wavelengths.  It was last reported to have
radio outbursts in 1975 at $962$ and $151$ MHz (Davis et al. 1975; Owen et al. 1976).
A few years after these observations, Duldig et al. (1979) reported a low level activity at
$2$ cm ($14.7$ GHz). More recent re-analysis of the $1975$ data revealed that it
underwent multiple jet ejection events (Kuulkers et al. 1999). There are 
no other reports of radio observations of this object. The outbursts and quiescence 
states are thought to be due to some form of thermal-viscous-instability in the accretion 
disk.  In the quiescent state, the accretion rate becomes very low (e.g. Lasota 2001).
Assuming that there is a Keplerian disk, from the observations in the optical and the 
X-ray, the accretion rate was estimated to vary from a few times the Eddington rate in 
outbursts to less than $ 10^{-11} M_\odot$ yr$^{-1}$ in quiescence 
(de Kool 1988; McClintock \& Remillard 1986).
Assuming a low-efficiency flow model, McClintock \& Remillard (2000), obtained 
the accretion rate to be $\sim 10^{-10} M_\odot$ yr$^{-1}$ using X-ray observations.  
A0620-00 has been in a quiescent state for quite some time. 
In the present Paper, we report the observation of a micro-flare in radio wavelength 
(frequency 1.28 GHz) coming from this object. 


\section{Observations and results}
\label{sect:Obs}
On Sept. 29th, 2002, during UT 00:45-02:03 we observed A0620-00 with the Giant Meter Radio
Telescope (GMRT) located in Pune, India. GMRT has $30$ parabolic reflector antennae placed with altazimuth
mounts each of which is of $45$ meter diameter placed in nearly `Y' shaped array. It has a tracking 
and pointing accuracy of $1'$ for wind speeds less than $20$ km/s. 
GMRT is capable of observing at six frequencies from $151$ MHz to $1420$ MHz.
On the higher side, $608-614$ MHz and $1400-1420$ MHz are protected frequency bands
by the International Telecommunication Union (ITU).
During our observation, 28 out of 30 antennae
were working and the observational conditions were stable. 
The observed frequency is $\nu_{obs} \sim 1280$ MHz which is far away from the ITU bands.
The band width is $16$ MHz. There were 128 channels with  
a channel separation of $125$ kHz.
We used 3C147 as the flux calibrator and 0521+166 as the phase calibrator.
No other source was found within the field of view. The primary beam width was 
$0.5$ degree and the synthesized beam width was $3$ arc second.
\begin{figure}
   \begin{center}
   \mbox{\epsfxsize=0.8\textwidth\epsfysize=0.8\textwidth\epsfbox{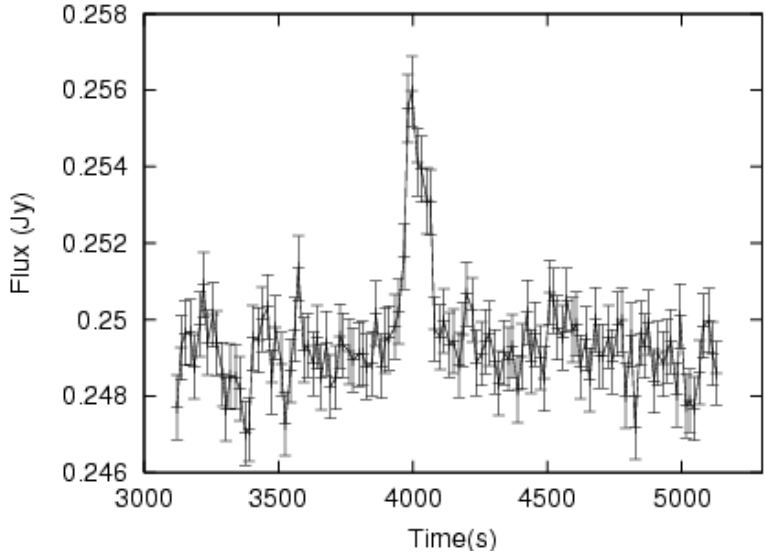}}
   \caption{Light curve of A0620-00 without background subtraction on Sep. 29th, 
2002 as obtained by GMRT radio observation at $1.28$ GHz. Subtracting 
the background reveals a micro-flare of mean flux $3.84$ mJy of duration $192 \pm 32 s$.}
   \end{center}
\end{figure}

Data analysis were carried out using the AIPS package. 
The data for A06200-00 is band passed and self calibrated.
The light-curve without the background subtraction is shown in Fig. 1.
The data is integrated in every $16$ seconds. The background is
due to two side lobes and is found to be constant in time.
The UV coverage was very good and the background was found to be constant
within the field of view with rms noise $8.6 \times 10^{-4}$ Jy as tested by the task IMAGR in AIPS.
The background subtraction reveals that a micro-flare
of average flux density $F_\nu=3.84$ mJy occurred and it lasted for
about $t_{\mu f} = 192 \pm 32$ seconds.
We found that each of the antennae independently showed this event
and the synthesized image of the field of view showed no significant signal 
from any other source. 
This confirms the presence of this micro-flare very convincingly.

\section{Interpretation of the micro-flaring event}
\label{sect:disc}
Fast variabilities occur in time scales of the order of the light
crossing time $t_l = r_g/c \sim 0.1\frac{M}{10M_\odot}$ ms, ($r_g=2GM/c^2$ is the
Schwarzschild radius) in the vicinity of a black hole. Shot noise in this time scale is
observed during X-ray observations. Since the duration $t_{\mu f}$ of the micro-flare
that we observe is much larger ($t_{\mu f}>>t_l$), hence we rule out 
the possibility that it is a shot noise type event. 
                                                                                
Assuming that the flare is due to magnetic dissipation, with an energy
density of $B^2/8\pi$, the expression for the total energy release (fluence) is:
$$
E_{mag} = \frac{B^2}{8\pi} V = 4 \pi D^2 \nu_{obs} F_\nu t_{\mu f} ,
\eqno{(1)}
$$
where $V \sim r_g^3$ is the lower limit of the volume in the accretion flow that released 
the energy, $D$ is the distance of the source from us, $\nu_{obs}$ is the frequency at which 
the observation is made and $F_\nu$ is the specific intensity of radiation. Here, $B$ is the
average magnetic field in the inflow where the flare forms. Re-writing Eqn. (1) using the 
equipartition law,
$$
\frac{B^2}{8\pi} \sim \frac{GM\rho}{r} =\frac{GM{\dot M}}{4 \pi v r^3} ,
\eqno{(2)}
$$
where $\rho$ is the density of matter in the accretion flow, ${\dot M}$ is 
the accretion rate and $v$ is the velocity of inflow.
Since there is no signature of a Keplerian disk in the quiescent 
state, one may assume the inflow to be generally like a low-angular momentum 
flow (Chakrabarti, 1990), especially close to the black hole. Estimations of 
McClintock \& Remillard (2000) was carried out with a low-efficiency radial 
flow model. Thus, we use the definition of the
accretion rate to be ${\dot M}=4\pi \rho r^2 v$. More specifically, we assume, 
the free-fall velocity,  $v \sim (2GM/r)^{1/2}$. Introduction of
pressure and rotation effects do not change the result since the gas is tenuous, i.e., hot,
and since the Keplerian flow is absent, i.e., the angular momentum is very low. These
simple but realistic assumptions allow us to obtain the upper limit of the
accretion rate of the flow to be,
$$
{\dot M} \sim (3.5 \pm 0.58) \times 10^{14} x^{5/2} {\rm \ gm/s}
= (5.5 \pm 0.91) \times 10^{-12} x^{5/2} M_\odot {\rm yr}^{-1}.
\eqno{(3)}
$$
Here $x=\frac{r}{r_g}$, is the dimensionless distance of the flaring
region from the center. From the transonic flow
models (Chakrabarti 1990), the flow is expected to be supersonic only
around $x_c\sim 2-3$ before disappearing into the black hole. Ideally, in a 
subsonic flow ($x>x_c$), the chance of flaring is higher as the residence time
of matter becomes larger, or comparable with the reconnection time scale. 
For $x<x_c$ there is little possibility of flaring. We thus estimate the
the accretion rate of A0620-00 in the quiescent state to be
$$
{\dot M}  =  (8.5 \pm 1.4) \times 10^{-11} (\frac{x}{3})^{5/2} M_\odot {\rm yr}^{-1}.
\eqno{(4)}
$$
In the case of a low angular momentum flow, there are possibilities of shock formation
at around $x\sim 10$ (Chakrabarti, 1990). So it is likely that the flare forms 
in the immediate vicinity of 
the post-shock (subsonic) region where the density of matter as well as magnetic pressure
are very high. In any case, the rate we get is consistent with that reported 
by McClintock \& Remillard (2000) on the basis of X-ray observations.
It is to be noted that Duldig et al. (1979) found a flux of $44\pm14$ mJy well after the
outburst in 1975 and concluded that intermittent emissions are possible and that mass transfer
continues even in quiescence states. Our result also verifies such an assertion. 
 
The procedure we have suggested here is sufficiently general.  For instance, it is 
generally believed that 
in the hard state of a black hole, the hot, sub-Keplerian matter plays an important role in
producing the so-called Compton cloud and this would an ideal location for flaring activities
if some entangled  magnetic fields are present.
In case the mass of the black hole and its distance are known, as in the present case, the
mass accretion rate could be calculated. In case the accretion rate and the distance were
known then the mass could be estimated by inverting the logical steps given above. One of 
our assumptions is to estimate the magnetic field by assuming it to be in equipartition
with the gravitational energy density. In reality, the magnetic field could be less
than the equipartition value. On the other hand, since we assumed flow to be freely 
falling, while in presence of angular momentum, the flow would slow down and both 
the density and the magnetic field energy would be higher. These to opposite effects 
should make our estimate to be still sufficiently realistic.

\acknowledgements We thank the staffs of the GMRT who have helped us to make this observation
possible. GMRT is run by the National Centre for Radio Astrophysics of the
Tata Institute of Fundamental Research. SP thanks a CSIR Fellowship which supported his
work at the Centre for Space Physics.

\label{lastpage}

\end{document}